\documentstyle{article}
\begin{document}

\begin{center}
{\LARGE \bf Photometry of the long period dwarf nova GY~Hya}\footnote{Based 
on observations taken at the Observat\'orio do Pico dos Dias / LNA and at 
CBA Pretoria}

\vspace{1cm}

{\Large \bf Albert Bruch}

\vspace{0.5cm}
Laborat\'orio Nacional de Astrof\'{i}sica, Rua Estados Unidos, 154,
CEP 37504-364, Itajub\'a - MG, Brazil

\vspace{0.5cm}
{\Large \bf Berto Monard}

\vspace{0.5cm}
Bronberg and Kleinkaroo Observatories, CBA Pretoria/Kleinkaroo,
Sint Helena 1B, PO Box 281, Calitzdorp 6660, South Africa

\vspace{0.8cm}

(Accepted for publication in New Astronomy)
\vspace{0.8cm}

{\bf Abstract}

\vspace{0.5cm}
\end{center}

\begin{abstract}
Although comparatively bright, the cataclysmic variable GY~Hya has not
attracted much attention in the past. As part of a project to better
characterize such systems photometrically, we observed light curves in
white light, each spanning
several hours, at Bronberg Observatory, South Africa, in 2004 and 2005,
and at the Observat\'orio do Pico dos Dias, Brazil, in 2014 and 2016. These
data permit to study orbital
modulations and their variations from season to season. The orbital
period, already known from spectroscopic observations of
Peters \& Thorstensen (2005), is confirmed through strong
ellipsoidal variations of the mass donor star in the system and the
presence of eclipses of both components. A refined period of
0.34723972~(6) days and revised ephemeries are derived. Seasonal changes in
the average orbital light curve can qualitatively be explained by 
variations of the contribution of a hot spot to the system light
together with changes of the disk radius. The amplitude of
the ellipsoidal variations and the eclipse contact phases permit to
put some constraints on the mass ratio, orbital inclination and the relative
brightness of the primary and secondary components. There are some indications
that the disk radius during quiescence, expressed in units of the component
separation, is smaller than in other dwarf novae.


\end{abstract}

\section{Introduction}
\label{Introduction}

Cataclysmic variables (CVs) are binary stars where a Roche-lobe filling
late-type component (the secondary) transfers matter via an accretion disk
to a white dwarf primary. It may be surprising that even after decades of
intense studies of CVs there are still an appreciable number of known or
suspected systems, bright enough to be easily observed with comparatively small
telescopes, which have not been studied sufficiently for basic parameters
to be known with certainty. In some cases even
their very class membership still requires confirmation.

Therefore, a small observing project aimed at a better understanding
of these stars was initiated in 2014 at the Observat\'orio do Pico dos
Dias (OPD), operated by the Laborat\'orio Nacional de Astrof\'{\i}sica, Brazil,
using 60cm class telescopes. First results have been published by
Bruch (2016, 2017) and Bruch \& Diaz (2017). Here, we present time
resolved photometry of the long period dwarf nova GY~Hya in order to
characterize orbital modulations.

Hoffmeister (1963) classified GY~Hya as either a RR~Lyr or U~Gem star.
Describing spectroscopic observations, Bond \& Tifft (1974) reject
the first alternative and consider the U~Gem classification as
probably correct. This is supported by the spectrum reproduced by
Zwitter \& Munari (1996) which shows the Balmer series strongly in emission.
He~I lines are also present. As an indication for the secondary star, the Na~D
lines are clearly visible in absorption. However, unlike normally seen in
quiescent dwarf novae, the He~II $\lambda$~4686~\AA\, emission line (and the
adjacent NIII-OIII complex) is particularly strong. This is more typically
the hallmark of magnetic cataclysmic variables, some novalike variables and
novae in quiescence (see e.g. Williams 1983). But
Cropper (1986) did not find the light of GY~Hya to be
significantly polarized.

The presence of the secondary star in the spectrum was confirmed by means
of time resolved spectroscopy performed by Peters \& Thorstensen (2005)
who measured the orbital period to be 0.347237 days. Quoting a private
communication from one of us (B. Monard) they also mention the presence
of eclipses with the same period in the light curve. This is based on data
analyzed in more detail in the present contribution.

This study is organized as follows:
In Sect.~\ref{Observations and data reductions} the observations and data
reduction techniques are briefly presented. Sect.~\ref{GY Hya} then deals
with the results of the observations which are subsequently discussed in
Sect.~\ref{Discussion}. Finally, a short summary
in Sect.~\ref{Summary} concludes this paper.

\section{Observations and data reductions}
\label{Observations and data reductions}

GY~Hya was observed at the Centre of Backyard Astronomy (CBA), Pretoria
(Bronberg Observatory), in 2002, 2004 and 2005, and
at OPD in 2014 and 2016.
The observations performed at OPD were obtained at the 0.6-m Zeiss and the
0.6-m Boller \& Chivens telescopes.
Time series imaging of the field around the target star was performed
using cameras of type Andor iKon-L936-B and iKon-L936-EX2 equipped with
back illuminated, visually optimized CCDs. Aiming to resolve
the expected rapid flickering variations the integration
times were kept short. Together with the small readout times of the detectors
this resulted in a time resolution of the order of 
$5^{\raisebox{.3ex}{\scriptsize s}}$. In order to
maximize the count rates in these short time intervals no filters were
used. The observations at CBA were also obtained 
in white light, using a Meade 30cm LX200 telescope equiped with a 
SBIG ST7 detector and focal reducer. The time resolution of the
light curves was $\sim$60 sec in 2002 and $\sim$30 sec in 2004 and 2005. 
A summary of the observations is
given in Table~\ref{Journal of observations}. Some light curve contain
gaps caused by intermittent clouds or technical reasons.

\begin{table}

\caption{Journal of observations of GY~Hya}
\label{Journal of observations}

\hspace{1ex}
\begin{tabular}{lccll}
\hline
Date & Start & End & Obs.* & \multicolumn{1}{l}{mean}  \\
     & (UT)  & (UT)&       & \multicolumn{1}{l}{mag**} \\
\hline
2002 Mar 06/07 & 23:31        & \phantom{2}3:21 & BO  & 15.3 \\
2002 Apr 20/21 & 17:45        & \phantom{2}1:31 & BO  & 15.3 \\
2002 Apr 22    & 16:51        & 19:33           & BO  & 15.3 \\
2002 May 09    & 17:20        & 19:59           & BO  & 15.3 \\
2002 May 27    & 19:12        & 22:45           & BO  & 15.3 \\ [1ex]
2004 May 13/14 & 16:29        & \phantom{2}1:18 & BO  & 14.5 \\
2004 May 22/23 & 16:16        & \phantom{2}1:11 & BO  & 15.3 \\
2004 Jun 04    & 18:22        & 23:26           & BO  & 15.3 \\
2004 Jun 06    & 16:23        & 19:06           & BO  & 15.4 \\ [1ex]
2005 Apr 11/12 & 21:26        & \phantom{2}3:31 & BO  & 15.2 \\ [1ex]
2014 May 01 & \phantom{2}0:46 & \phantom{2}1:29 & OPD & 15.8 \\
2014 May 03 & \phantom{2}3:26 & \phantom{2}4:50 & OPD & 15.8 \\
2014 Jun 19/20 & 21:17        & \phantom{2}2:07 & OPD & 15.8 \\
2014 Jun 21/22 & 21:38        & \phantom{2}2:07 & OPD & 15.9 \\ [1ex]
2016 Apr 06 & \phantom{2}6:13 & \phantom{2}8:27 & OPD & 15.8 \\
2016 Arp 07 & \phantom{2}6:43 & \phantom{2}8:31 & OPD & 15.9 \\
2016 Apr 08 & \phantom{2}6:18 & \phantom{2}8:32 & OPD & 15.9 \\
2016 May 10 & \phantom{2}2:03 & \phantom{2}4:10 & OPD & 16.0 \\
2016 May 11/12 & 22:09        & \phantom{2}5:32 & OPD & 15.8 \\
2016 May 12/13 & 21:34        & \phantom{2}0:15 & OPD & 15.9 \\
2016 Jun 09 & \phantom{2}0:26 & \phantom{2}3:54 & OPD & 15.8 \\
2016 Jun 28/29 & 21:10        & \phantom{2}0:56 & OPD & 15.8 \\ [1ex]
\hline
\multicolumn{5}{l}{\phantom{*}*BO\phantom{D} = Bronberg Observatory}\\
\multicolumn{5}{l}{\phantom{**}OPD = Observat\'orio do Pico dos Dias}\\
\multicolumn{5}{l}{** $R$ for BO; $V$ for OPD}\\
\end{tabular}
\end{table}
%

Since all observations were performed in white light, it was not possible to
calibrate the stellar magnitudes. Instead, the brightness was measured as
the magnitude difference between the target and the nearby comparison star.
For the OPD data, UCAC4 321-074490 ($V = 11.901$; Zacharias et al. 1993)
was chosen as comparison star. 
For the CBA data it was UCAC4 321-074480 ($R = 13.3$)
Their constancy was verified through the observations of check
stars. A rough estimate of the effective
wavelength of the white light band pass, assuming a mean atmospheric extinction
curve, a flat transmission curve for the telescope, and a detector efficiency
curve as provided by the manufacturer, yields $\lambda_{\rm eff} \approx
5530$~\AA\ for the OPD instrumentation, very close to the effective wavelength 
of the Johnson $V$ band (5500~\AA; Allen 1973). The detector
sensitivity of the CCD used at CBA leads to an effective
wavelength close to that of the $R$ band.
Therefore, using the known magnitudes of the comparison star it is possible to
calculate approximate mean nightly magnitudes of the target star.
These are included in Table~\ref{Journal of observations}.
Assuming that the quiescent brightness of GY~Hya did not change significantly
the difference of 
$\approx$$0^{\raisebox{.3ex}{\scriptsize m}}_{\raisebox{.6ex}{\hspace{.17em}.}}5$
between the OPD $V$ band and the
CBA $R$ band observations is quite normal for a CV with
a dominant contribution from the secondary star (see Sect.~\ref{Discussion}).

Data reduction of the CBA data and the construction of light 
curves was done using AIP dataprocessing software. Basic data reduction 
(biasing, flat-fielding) of the OPD data was performed using IRAF. Light 
curves were derived with aperture photometry routines implemented in the 
MIRA software system (Bruch 1993). The latter
system was also used for all further data reductions and calculations.
Throughout this
paper time is expressed in UT. However, whenever observations taken in
different nights were combined (e.g., to search for orbital variations)
time was transformed into barycentric Julian Date on the Barycentric
Dynamical Time (TDB) scale using the online tool provided by
Eastman et al.\ (2010) in order to take into account variations of the
light travel time within the solar system.

\section{Results}
\label{GY Hya}

\begin{figure}
   \parbox[]{0.1cm}{\epsfxsize=14cm\epsfbox{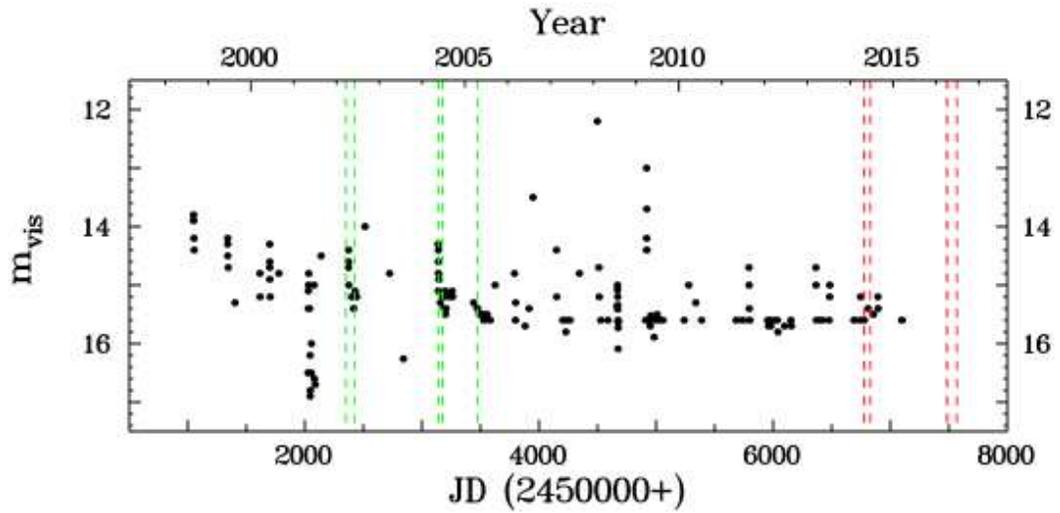}}
      \caption[]{Long term light curve of GY~Hya based on AAVSO data.
                 The broken lines delimit the epochs of
                 the time resolved CBA (green) and
                 OPD (red) observations. (For interpretation
                 of the references to colour in this figure legend, the
                 reader is referred to the web version of this article.)}
\label{gyhya-aavso}
\end{figure}

The long term light curve of GY~Hya based on data of the American
Association of Variable Star Observers (AAVSO) is shown in
Fig.~\ref{gyhya-aavso} (one low data point marked as uncertain has been
rejected). Several high points may be interpreted as
outbursts and thus corroborate the spectroscopic classification of the
system as a dwarf nova. On the other hand GY~Hya also occasionally drops
into a state of low brightness (but see the discussion in
Sect.~\ref{Discussion}). Interestingly, the system magnitude
(disregarding outbursts and low states) seems to decline steadily from
$\sim$$15^{\raisebox{.3ex}{\scriptsize m}}$ to 
$\sim$$15^{\raisebox{.3ex}{\scriptsize m}}_{\raisebox{.6ex}{\hspace{.17em}.}}6$
between 1998 and 2005 and remains stable thereafter.

\input epsf

\subsection{The orbital light curve}
\label{ GY Hya The orbital light curve}

\begin{figure}
   \parbox[]{0.1cm}{\epsfxsize=14cm\epsfbox{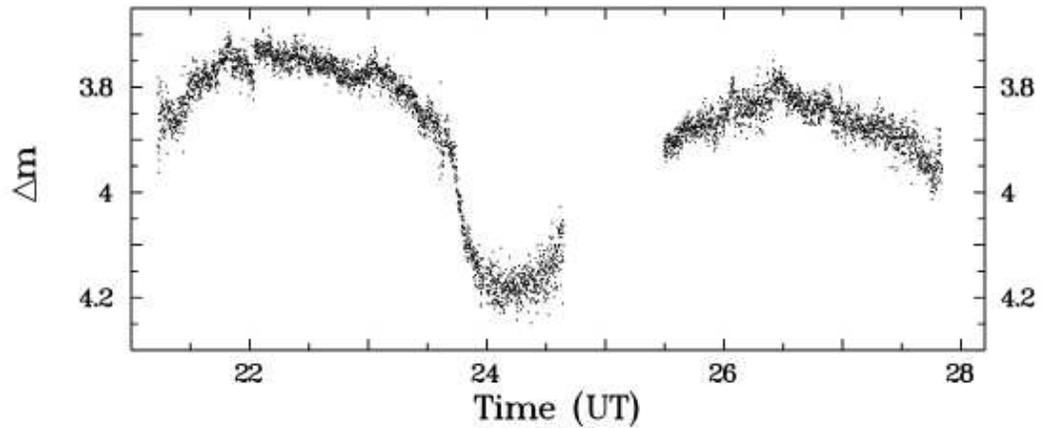}}
      \caption[]{Light curve of GY~Hya of 2016, June 9/10.}
\label{gyhya-lightc}
\end{figure}

The epochs of our time resolved observations are delimited by broken
red (OPD data) and green (CBA data) vertical lines
in Fig.~\ref{gyhya-aavso}. 
As an example, the data of 2016, June 9/10 are shown in
Fig.~\ref{gyhya-lightc}. Pronounced variations on hourly time scales are
seen, interrupted by an eclipse (the egress of which only being partially
observed due to intermittent clouds). Flickering is clearly present, but
only on a low level, hardly standing out against the data noise.

The long orbital period makes it difficult to observe a continuous light
curve covering all phases. However, stitching together data from various
nights it is possible to construct a mean orbital light curve. In order to
take into account possible night-to-night variations,
in an interactive process the mean
magnitude difference between the phase folded nightly data and the
respective phase range of the average orbital light curve was calculated
and subtracted.

\begin{figure}
   \parbox[]{0.1cm}{\epsfxsize=14cm\epsfbox{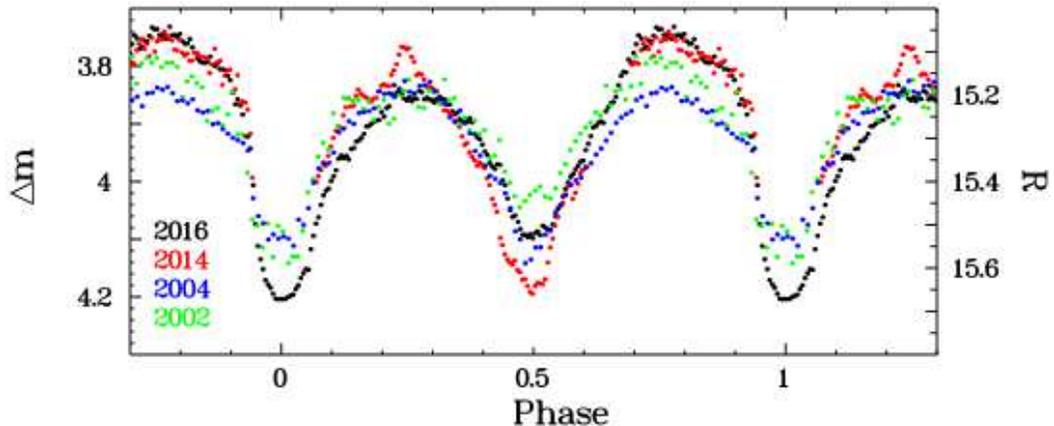}}
      \caption[]{Average orbital light curves of GY~Hya in different years. The
                 left scale refers to the differential magnitudes of 2014
                 and 2016, the right one to those of
                 2002 and 2004, expressed as approximate $R$ magnitudes.
                 (For visualization of the colours used in this figure, the
                 reader is referred to the web version of this article.)}
\label{gyhya-fold}
\end{figure}

Mean orbital light curves were constructed from the observations during
each year. The corresponding time base spans between 0.5 and 3 month.
The mean curves are shown in
Fig.~\ref{gyhya-fold}, binned in phase intervals of 0.01 (2002 and 2004)
and 0.005 (2014 and 2016), respectively. Mid-eclipse according to the
revised ephemeris (see Sect.~\ref{Ephemeris}) has been chosen as phase 0.
Since in 2005 GY~Hya was observed
only in one night, not covering all phases, these data are not regarded here.
Moreover, the light curve on 2004, May 13 was not used because the star was
in a brighter state at this epoch
(see Sect.~\ref{GY Hya Outburst light curve}).

In general terms, the orbital light curve is characterized by two strongly
expressed minima and maxima. The secondary minimum close to phase 0.5 is
-- depending on the epoch -- almost as deep as the primary minimum,
while the maximum after the primary minimum is slightly fainter than
that after the secondary minimum. However, there are differences from
epoch to epoch. This is most evident comparing the year 2004 (blue curve
in Fig.~\ref{gyhya-fold}), when both minima were of equal depth and both
maxima of equal height, with 2016 (black curve), when the strongest
differences of both, the minima and the maxima, are observed and the
primary minimum is deepest. In 2014 the secondary minimum reached almost
the depth of the primary minimum in 2016. Unfortunately, the primary
minimum was not covered in that year.

\subsection{Ephemeris}
\label{Ephemeris}

The spectroscopic period measured by Peters \& Thorstensen (2005) and
the photometric period quoted by them rely on a comparatively short time
base. Combining the new observations of 2014 and 2016 with the those of
2002 -- 2005 enlarges significantly the time base for the precise
determination of the orbital period. Instead of measuring individual eclipse
times a representative minimum epoch for each observing season was
derived from the average orbital light curves of each year. Only for
2005, with just one partial light curve available, the minimum epoch
was measured directly. To best determine the location of the bottom
of the eclipse, polynomials of various degrees (between 2 -- 5) and a
Gaussian were fitted to the eclipse profile\footnote{For the 2016 data
the polymial of degree 2 and the Gaussian clearly did not result in an
acceptable fit and therefore were disregarded.} and the epoch of the
respective extrema calculated. Their mean values together with their
standard deviations are listed in Table~\ref{GY Hya eclipse epochs}.
A linear least squares fit to these data yields the updated ephemeris:

\begin{displaymath}
{\rm BJD_{min}} = 2457484.2337 (7) + 0.34723972 (6) \times E
\end{displaymath}

Here $E$ is the cycle number. The quoted uncertainties are formal fit errors
which may underestimate the true error. The distribution of data
points within two restricted time intervals separated by more than a decade
does not allow to determine a possible period derivative. Therefore, only
a linear ephemeries, i.e., the average period over the total time base, is
given.

\begin{table}

\caption{GY~Hya eclipse epochs}
\label{GY Hya eclipse epochs}

\hspace{1ex}

\begin{tabular}{lcc}
\hline
Year      & Cycle  No.         & Eclipse epoch  \\
          &                    & BJD (TDB)      \\
\hline
2002      & -14\,785           & 2\,452\,350.2959 $\pm$ 0.0018 \\
2004      & -12\,487           & 2\,453\,148.2508 $\pm$ 0.0018 \\
2005      & -11\,554           & 2\,453\,472.2248 $\pm$ 0.0007 \\
2016      & \phantom{-12\,48}0 & 2\,457\,484.2340 $\pm$ 0.0007 \\
\hline
\end{tabular}
\end{table}
%

\subsection{Outburst light curve}
\label{GY Hya Outburst light curve}

\begin{figure}
   \parbox[]{0.1cm}{\epsfxsize=14cm\epsfbox{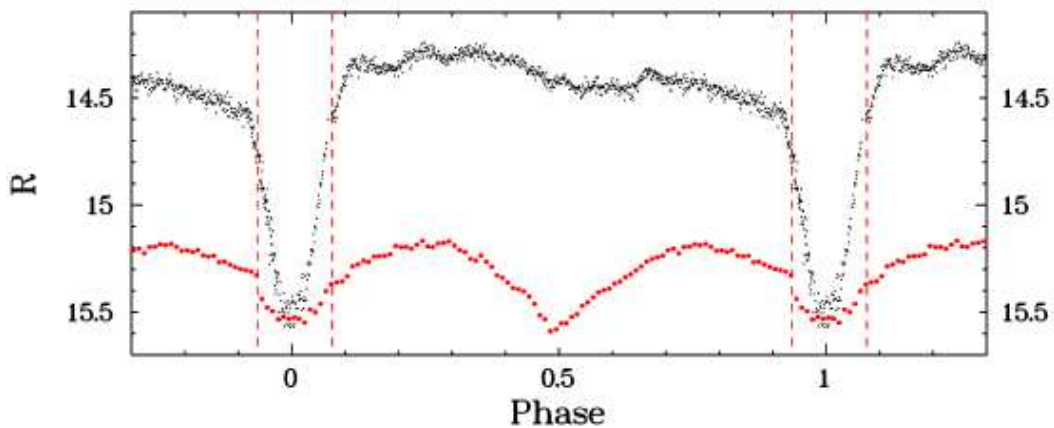}}
      \caption[]{Light curve of GY~Hya of 2004, May 13/14, observed during
                 an outburst, folded on the orbital period (black) together
                 with the average quiescent orbital light curve of 2004
                 (red). (For interpretation
                 of the references to colour in this figure legend, the
                 reader is referred to the web version of this article.)}
\label{gyhya-outburstlc}
\end{figure}

On 2004, May 13/14 GY~Hya was observed to be 
$0^{\raisebox{.3ex}{\scriptsize m}}_{\raisebox{.6ex}{\hspace{.17em}.}}8$ brighter
than normal. The long-term AAVSO light curves shows that it has been at
about the same brightness level for at least 3 days around this date.
Nine days later the normal quiescent level is reached again. Although the
sparse data do not permit to say if the late phase of a genuine dwarf nova
outburst has been observed, for simplicity we will subsequently refer to
this high state as an outburst.
The shape of the outburst orbital light curve, shown in black in
Fig.~\ref{gyhya-outburstlc}, is drastically different from the
average (quiescent state) curve during the other observing nights
of the same year, shown in red in the figure. While the bottom of the
primary minimum (the eclipse)
attains the same magnitude in both states the amplitude of the eclipse
is much larger during outburst. The width of the eclipse in both states,
indicated by the broken vertical lines in the figure, is very similar
($\Delta \phi = 0.140$); at most the eclipse may start slightly earlier
during outburst. It is ``V''-shaped in outburst but flat bottomed during
quiescence.

The secondary minimum near phase 0.5 is still apparent during outburst,
but it is much shallower. We measured its depth after transforming
the light curves from magnitudes to intensities (on an arbitrary scale).
During quiescence, its amplitude is taken to be the difference between
the maxima on either side and the minimum. During outburst, when GY~Hya
does not attain the same light level after the secondary minimum that it
had before, a third order polynomial fit to the the phase ranges 0.12 --
0.41 and 0.7 -- 0.89 was first subtracted before determining the
intensity difference. The amplitude of the secondary minimum during
outburst is then found to be 36\% of the amplitude during quiescence.

\section{Discussion}
\label{Discussion}

At the long orbital period of 
$8^{\raisebox{.3ex}{\scriptsize h}}$~$20^{\raisebox{.3ex}{\scriptsize m}}$
the secondary star of
GH~Hya is expected to contribute significantly to the total light; more
than the relatively faint absorption lines in the spectrum shown by
Zwitter \& Munari (1996) would suggest. But then, their observations were
made when the system was at 
$V = 15^{\raisebox{.3ex}{\scriptsize m}}_{\raisebox{.6ex}{\hspace{.17em}.}}08$, more than 
half a magnitude above its normal quiescent brightness. On the other hand, 
during the observations of Peters \& Thorstensen (2005) the spectrum was
dominated by strong absorption lines of spectral type K4 or K5.

The orbital variations of GY~Hya documented here confirm that much of the
system light in the optical band comes from the secondary star. They can
be interpreted as ellipsoidal variations of the mass donor in GY~Hya
in combination with eclipses of both system components. As is evident from
Fig.~\ref{gyhya-fold}, the shape of the variations changes from season to
season. While the maximum at phase 0.25 remains fairly stable this is not the 
case for the second maximum. Qualitatively, this can be explained by a variable
contribution of a hot spot which hides behind the disk at phases close to
0.25 but is in plain sight during the second maximum. Such variations may
either be caused by a real time dependence of its brightness, or by a 
variable contribution of the hot spot in the different photometric bands,
or both. Moreover, the depth of the minima (eclipses) changes considerably.
Since this is seen in both, the OPD and the CBA data, this cannot
be explained as being due to the different passbands. Instead, disk radius 
variations may be invoked. If the eclipses are
partial -- as is strongly suggested by their shape -- the more concentrated
light from a smaller disk would lead to deeper eclipses at primary minimum
(phase 0, when the secondary star is in front) unless only
the outer parts of the disk are eclipsed. The secondary minimum should
be shallower because a smaller part of the mass donor is eclipsed.

\begin{figure}
   \parbox[]{0.1cm}{\epsfxsize=14cm\epsfbox{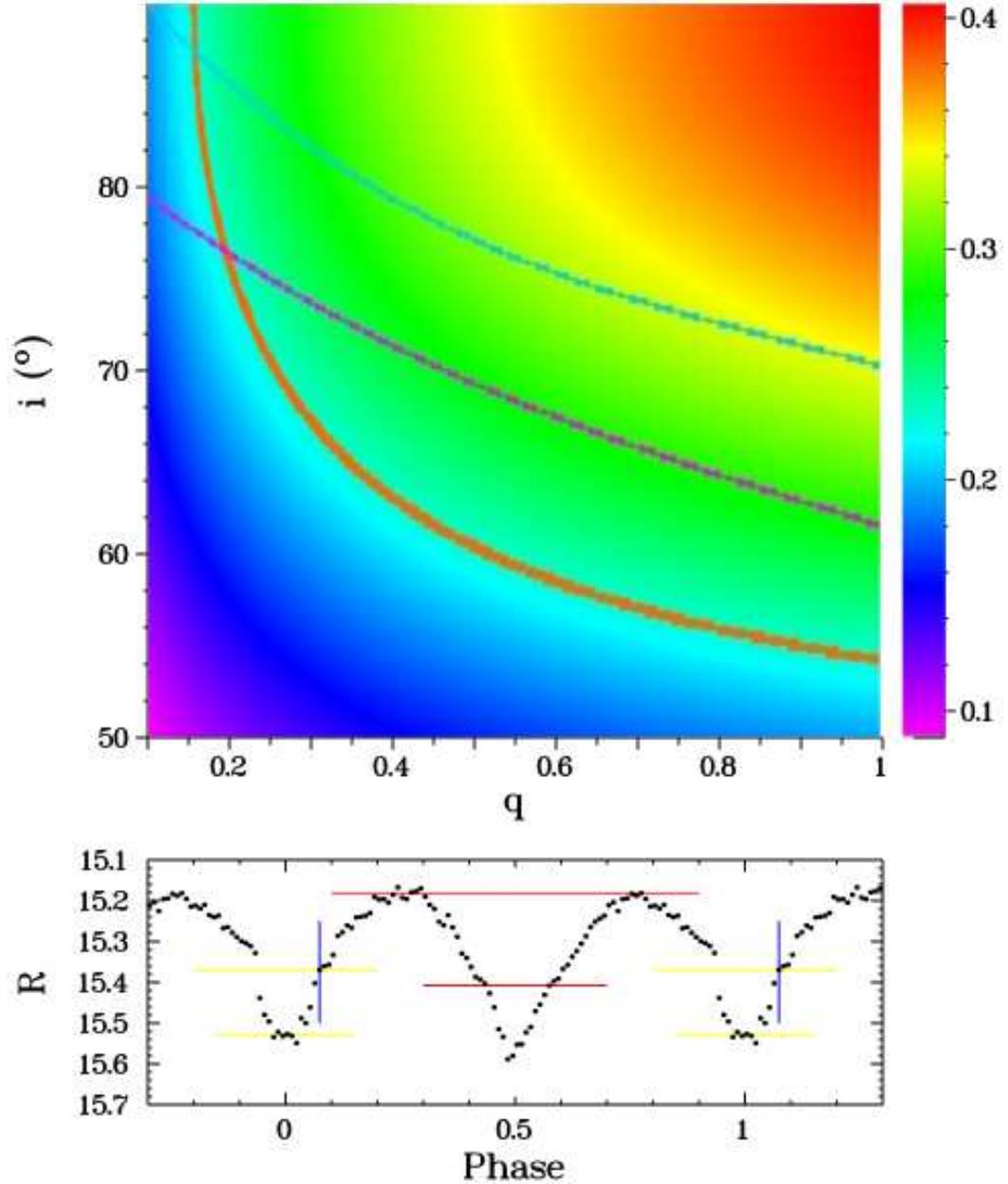}}
      \caption[]{{\it Bottom:} Mean phase folded light curve of GY~Hya observed
                  in 2004. The magnitudes levels of the orbital maxima and of
                  ingress and egress of the secondary eclipse are marked by
                  red horizontal lines. Their difference $\delta m$ is a lower
                  limit to the ellipsoidal variation of the secondary star.
                  Similarly, the difference between the yellow horizontal
                  lines defines an upper limit to the primary eclipse depth
                  $\Delta m_{\rm ecl}$. The blue vertical lines indicate the
                  outer contact phase of the primary eclipse. {\it Top:} The
                  total amplitude $\Delta m$ of ellipsoidal variations
                  (the colour code is specified
                  on the colour bar at the right) of the
                  secondary star in GY~Hya as a function of the assumed
                  mass ratio $q$ and orbital inclination $i$. The brown solid
                  curve separates permitted $q - i$ combinations ($\Delta m >
                  \delta m$) to the upper right from forbidden combinations
                  ($\Delta m < \delta m$) to the lower left. The light blue and
                  violet solid lines represent the functional relationship
                  between $q$ and $i$ derived from the observed outer eclipse
                  contact phase and Roche geometry for assumed disk radii of
                  $0.2 A$ and $0.3 A$, respectively, where $A$ is the distance
                  between the primary and secondary components. (For
                  interpretation
                  of the references to colour in this figure legend, the
                  reader is referred to the web version of this article.)}
\label{gyhya-ellip}
\end{figure}

The ellipsoidal variations of GY~Hya provide a handle to put (albeit only
weak) constrains on the mass ratio $q$ and the orbital inclination $i$.
The Wilson-Devinney (Wilson \& Devinney 1971, Wilson 1979) code was used to
calculate $R$ band light curves
of a Roche-lobe filling star with a temperature of 4130~K, appropriate for
a K5 main sequence star (Allen 1973), as a function of $q$ and $i$,
fixing the limb and gravity darkening coefficients to values interpolated
in the tables of Claret \& Bloemen (2011). The albedo is irrelevant because no
illumination of the star is assumed. The total amplitude $\Delta m$ of the
resulting ellipsoidal variations is plotted in Fig.~\ref{gyhya-ellip} (upper
frame). It is colour coded as specified on the colour bar at the right of the
figure. The 2004 light curve of GY~Hya, being the most symmetrical of all
with almost
equal maxima, appears to be least distorted by variable contributions of
other light sources around the orbit. For convenience, it is included in
Fig.~\ref{gyhya-ellip} (lower frame).
The onset and end of the secondary eclipse
(phase 0.5) can be clearly identified. The corresponding magnitude level is
marked by a red horizontal line in the figure. The second horizontal line marks
the maximum magnitude level. Taking into account that the contribution of
other light sources to the total brightness will reduce the observed amplitude,
the difference 
$\delta m = 0^{\raisebox{.3ex}{\scriptsize m}}_{\raisebox{.6ex}{\hspace{.17em}.}}226$
between these lines is then a
firm lower limit for $\Delta m$ of the ellipsoidal
variations of the secondary star. It is shown as a brown curve in the
upper frame of the figure which separates the permitted $q - i$ combinations
in the upper right part from combinations which would lead to too small an
amplitude for the ellipsoidal variations.

Assuming the radius $R_{\rm d}$ of the accretion disk (taken to be flat
and infinitesimally thin) to be known, Roche geometry and the last contact
phase of the disk eclipse provide a functional relationship between $q$ and
$i$. The last contact is quite well defined in the light curve. It occurs at
phase 0.074 (blue vertical lines in Fig.~\ref{gyhya-ellip}, lower frame). In
quiescent dwarf novae $R_{\rm d} \approx 0.3$ [in units of the component
separation; e.g.\ OY~Car: 0.31 (Wood et al.\ 1989); Z~Cha: 0.33 
(Wood et al.\ 1986); V2051~Oph: 0.32 (Baptista et al.\ 2007); 
V4104~Sgr: 0.28 (Borges \& Baptista 2005)]. Adopting
this value we numerically derived the $q - i$ relationship which is shown as a
violet line in the upper frame of the figure. 

For all values of $q > 0.3$, the
observed outer contact phase together with $R_{\rm d} = 0.3$ leads to a
configuration such that the central parts
of the disk remain uneclipsed. For $q \rightarrow 1$ only a small part of
the outer disk is invisible at mid-eclipse. This is hardly compatible with
the observed eclipse depth. Moreover, the eclipse depth during outburst
(Sect.~\ref{GY Hya Outburst light curve}) should then not reach the same low
level seen in quiescence because the central parts of the then bright 
accretion disk would remain visible. Therefore, in order to assess the effect
of a smaller disk radius, we also calculated the $q - i$ relation for
$R_{\rm d} = 0.2$ which is shown as a light blue line in the figure. We remark
that for this disk radius the centre of the primary is eclipsed for all $q$
and the disk eclipse is total for $q < 0.75$.

\begin{figure}
   \parbox[]{0.1cm}{\epsfxsize=14cm\epsfbox{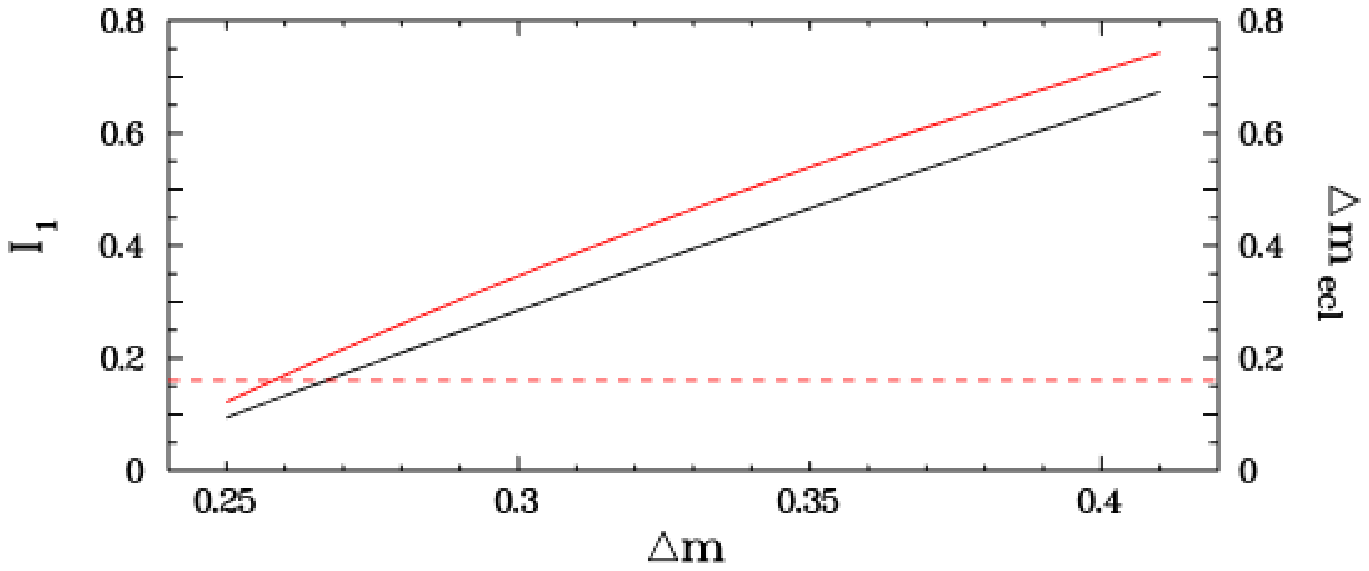}}
      \caption[]{Upper limit to the relative contribution $I_1$
                 of the primary star light of GY~Hya
                 (in units of the secondary star light at orbital maximum)
                 as a function of the intrinsic amplitude $\Delta m$ of the
                 ellipsoidal
                 variations of the secondary (black line, left hand scale).
                 Upper limit of the primary eclipse depth $\Delta m_{\rm ecl}$
                 (total eclipse assumed) as a function of $\Delta m$ (red line,
                 right hand scale). The broken line indicates the observed
                 eclipse depth. (For interpretation
                 of the references to colour in this figure legend, the
                 reader is referred to the web version of this article.)}
\label{gyhya-ecldepth}
\end{figure}

The strong ellipsoidal variations attest to the dominant contribution of the
secondary star to the total light. An upper limit to the relative contribution
of the primary component ($I_1$, expressed in units of the secondary
contribution $I_2$ at orbital maximum) can be calculated as a function of the
intrinsic amplitude of the ellipsoidal variations ($\Delta m$) from 
their observed amplitude ($\delta m$).
For $\Delta m$ between the maximum value in Fig.~\ref{gyhya-ellip} and a
value just above its permitted minimum value $\delta m$ the contribution
of the primary must remain below the back line in Fig.~\ref{gyhya-ecldepth}
(left hand scale).

Is this compatible with the observed primary eclipse depth $\Delta m_{\rm ecl}$?
Yes, it is.
Assuming the eclipse to be total, an upper limit for $\Delta m_{\rm ecl}$
can be calculated from $I_2$, $\Delta m$ and the upper limit to $I_1$.
This is shown as a red line in Fig.~\ref{gyhya-ecldepth} (right hand scale).
The observed primary eclipse depth can be estimated from the magnitude
difference at mid-eclipse and the end of eclipse egress in the lower frame of
Fig.~\ref{gyhya-ellip} to be not more than
$\Delta m_{\rm ecl} = 0^{\raisebox{.3ex}{\scriptsize m}}_{\raisebox{.6ex}{\hspace{.17em}.}}16$
(difference between the yellow
horizontal lines). This is an upper limit because due to
its ellipsoidal variations the secondary contributes less at mid-eclipse than
at the end of egress. $\Delta m_{\rm ecl}$ is shown as a broken line in
Fig.~\ref{gyhya-ecldepth}. Obviously, the observed eclipse depth is
significantly smaller than than the calculated upper limit for all but
the smallest values of $\Delta m$.

While the above considerations lead to a consistent picture of GY~Hya a
problem arises when regarding some low points which appear in the long
term light curve (Fig.~\ref{gyhya-aavso}). Even for the maximum upper limit of
$I_1$ (referring to the maximum of value of $\Delta m$ considered here) the
brightness of the system cannot decrease by more than 
$\sim$$0^{\raisebox{.3ex}{\scriptsize m}}_{\raisebox{.6ex}{\hspace{.17em}.}}56$
even under the extreme assumption that the primary star contribution drops
to zero. How can GY~Hya then attain
a magnitude as faint as 
$16^{\raisebox{.3ex}{\scriptsize m}}_{\raisebox{.6ex}{\hspace{.17em}.}}9$? 
Most of the low points are
concentrated in a small time interval between 2001, April 27 and June 27
and were observed by the same observer (who did not contribute any other
data). Moreover, at least on two occasions other observers found the system to
be at a ``normal'' magnitude just two days before or after a faint observation.
The one remaining low point at JD 2452842 is the only data point provided by
another observer. This may cast doubts on the reliability of the faint
data points.

\section{Summary}
\label{Summary}

We present the first time resolved photometry of the CV
GY~Hya. Its overall behaviour is as expected from a dwarf nova with a
long orbital period. The light curve is characterized by strong orbital
modulations in quiescence caused by ellipsoidal variations of the dominant
mass donor star together with eclipses of both system components.
As is normal in cataclysmic variables, flickering is present
but on a magnitude scale significantly smaller than in most quiescent
dwarf novae (Beckemper 1995) because the flickering light source,
expected to be associated
to the accretion disk, is outshone by the brighter secondary star. The
average light curves show
some variability in different observing seasons which can qualitatively be
explained by a variable contribution of a hot spot to the total light
together with changes of the disk radius. There are some indications that
the disk radius, expressed in units of the component separation, is smaller
than observed in other dwarf novae.

\section*{Acknowledgements}

We gratefully acknowledge the use of observations from the AAVSO International
Database contributed by observers worldwide. They provided valuable supportive
information for this study.

\section*{References}

\begin{description}

\item Allen, C.W. 1973, Astrophysical Quantities, third edition
      (Athlone Press: London)
\item Baptista, R., Santos, R.F., Fa\'undez-Abans, M., \& Bortoletto, A.
      2007, AJ, 134, 867
\item Beckemper, S. 1995, Statistische Untersuchungen zur St\"arke des
      Flickering in kataklysmischen Ver\"anderlichen,
      Diploma thesis, M\"unster
\item Bond, H.E., \& Tifft, W.G. 1974, PASP, 86, 981
\item Borges, B., \& Baptista, R. 2005, A\&A 437, 325
\item Bruch, A. 1993,
      MIRA: A Reference Guide (Astron.\ Inst.\ Univ.\ M\"unster)
\item Bruch, A. 2016, New Astr., 46, 90
\item Bruch, A. 2017, New Astr., 52, 112
\item Bruch, A., \& Diaz, M.P. 2017, New Astr., 50, 109
\item Claret, A., \& Bloemen, S. 2011, A\&A, 529, A75
\item Cropper, M. 1986, MNRAS 222, 225
\item Eastman, J., Siverd, R., \& Gaudi, B.S. 2010, PASP, 122, 935
\item Hoffmeister, C. 1963, Ver\"off.\ Sternw.\ Sonneberg, 6, 38
\item Peters, C.S., \& Thorstensen, R. 2005, PASP 117, 1386
\item Williams, G. 1983, ApJ Suppl., 53, 523
\item Wilson, R.E. 1979, ApJ, 234, 1054
\item Wilson, R.E., \& Devinney, E.J. 1971, ApJ, 166, 605
\item Wood, J.A., Horne, K., \& Berriman G. 1989, ApJ, 341, 974
\item Wood, J.A., Horne, K., Berriman G., et al. 1986, MNRAS, 219, 629
\item Zacharias, N., Finch, C.T., Girard, T.M., et al.\ 2013, AJ, 145, 44
\item Zwitter, T., \& Munari, U. 1996, A\&AS, 117, 449

\end{description}

\end{document}